\documentclass[conference,10p]{IEEEtran}
\usepackage{tikz}
\usepackage{textcomp}

\newcommand\copyrighttext{%
  \footnotesize \textcopyright 2022 IEEE. Personal use of this material is permitted.
  Permission from IEEE must be obtained for all other uses, in any current or future
  media, including reprinting/republishing this material for advertising or promotional
  purposes, creating new collective works, for resale or redistribution to servers or
  lists, or reuse of any copyrighted component of this work in other works.}
\newcommand\copyrightnotice{%
\begin{tikzpicture}[remember picture,overlay]
\node[anchor=south,yshift=10pt] at (current page.south) {\fbox{\parbox{\dimexpr\textwidth-\fboxsep-\fboxrule\relax}{\copyrighttext}}};
\end{tikzpicture}%
}
\usepackage[numbers]{natbib}
\usepackage{amsmath,amssymb,amsfonts}
\usepackage{algorithmic}
\usepackage{graphicx}
\usepackage{textcomp}
\usepackage{xcolor}
\usepackage{url}
\usepackage{balance}

\def\BibTeX{{\rm B\kern-.05em{\sc i\kern-.025em b}\kern-.08em
    T\kern-.1667em\lower.7ex\hbox{E}\kern-.125emX}}

\begin{document}

\title{IoTreeplay: Synchronous Distributed Traffic Replay\\in IoT Environments}

\author{\IEEEauthorblockN{Markus Toll}
\IEEEauthorblockA{\textit{Technische Universität Berlin} \\
Berlin, Germany \\
markus.toll@campus.tu-berlin.de}

\and

\IEEEauthorblockN{Ilja Behnke}
\IEEEauthorblockA{\textit{Technische Universität Berlin} \\
Berlin, Germany \\
i.behnke@tu-berlin.de}

\and

\IEEEauthorblockN{Odej Kao}
\IEEEauthorblockA{\textit{Technische Universität Berlin} \\
Berlin, Germany \\
odej.kao@tu-berlin.de}
}

\maketitle
\copyrightnotice

\begin{abstract}
Use-cases in the Internet of Things (IoT) typically involve a high number of interconnected, heterogeneous devices. Due to the criticality of many IoT scenarios, systems and applications need to be tested thoroughly before rollout. Existing staging environments and testing frameworks are able to emulate network properties but fail to deliver actual network-wide traffic control to test systems application independently. To extend existing frameworks, we present the distributed traffic replaying tool \emph{IoTreeplay}.

The tool embeds \textit{TCPLivePlay} into an environment that allows the synchronous replaying of network traffic with multiple endpoints and connections. Replaying takes place in a user-defined network or testbed containing IoT use-cases. Network traffic can be captured and compared to the original trace to evaluate accuracy and reliability. The resulting implementation is able to accurately replay connections within a maximum transmission rate but struggles with deviations from regular TCP connections, like packet loss or connection reset. An evaluation has been performed, measuring individual and aggregated delays between packets, based on the recorded timestamps.
\end{abstract}


\section{Introduction}
\label{sec:introduction}

The Internet of Things (IoT) has steadily grown over the last years. The development of more sophisticated sensor devices combined with new technologies, especially in areas of communication, like 5G, has multiplied potential applications. Multiple devices can be connected to interact autonomously based on self-created, shared, and evaluated data~\cite{Markettrends9210375}.
Household automation and DIY projects aren't the only fields of application for this kind of technology with the prefix "smart" being added to every aspect of life. Use cases include but are not limited to infrastructure applications, industrial use cases in fabrication and logistics processes, or improvements in the mobility sector~\cite{actualtraffic10.1007/978-3-030-51974-2_50, smartgrid, lasi2014industry}.
An increase in the number of devices and their usage and connectivity brings with it a massive increase in network traffic \cite{farooq2015review}.
IoT contributes increasingly to the sum of all network traffic and is being used in an ever increasing amount of public and business applications. Resulting from this is a demand for solutions that can guarantee safety, efficiency, and scalability, as well as enable testing of new and existing technologies.

Distributed IoT networks are unique in the aspect that they usually consist of a high number of heterogeneous devices. Each of these devices is characterized by different properties but all of them connected to the same network and possibly the internet. Next to the incidental scalability requirements~\cite{gomes2014internet}, IoT scenarios in critical infrastructures and the industry create a hard requirement for safety, namely reliability and predictable behavior~\cite{moore2020iot}. To this end, the used systems need to be tested thoroughly with scalability and realism in mind. Since there is no single point of control, it proves difficult to estimate how individual devices and the network as a whole would react to a change in network traffic. Examples of this could be the increase in data size, a decrease in breaks between periodic processes, or more network-centered changes like the introduction of delay, limitations of bandwidth, and more \cite{SRINIDHI20191}.
Additionally, a spectrum of protocols exists and is used to transport data depending on use case and application \cite{iotprotocols}.

To assist the development of IoT applications, several frameworks and staging environments exist~\cite{symeonides2020fogify, behnke2019hector, beilharz2021towards}. Additionally, a comprehensive review of the most prominent solutions has been performed~\cite{beilharzcontinuously}. It shows, that while some testing frameworks offer fault injection and network emulation tools they lack application independent network testing and the integration of domain specific environment simulations, which poses a problem since IoT networks are often characterised by the tight coupling and interaction within their environment.  

Within this work, we present \emph{IoTreePlay}, a tool that allows the synchronized distributed replaying of TCP traffic inside an entire network. It augments the capabilities of the \textit{TCPLivePlay} tool, which is part of the \textit{TCPReplay} suite\footnote{\url{https://github.com/appneta/tcpreplay}}. The goal is to enable a user to distribute previously recorded traffic onto multiple devices and replay them according to their original schedule as well as to record said traffic to evaluate it later. Traffic traces can be used on any IoT testbed containing real and/or virtual devices. Therefore, \emph{IoTreePlay} makes it possible to perform network testing independent of the environment in which the traffic trace was recorded. With its focus on TCP it provides capability to replay multiple applicaton-level protocols like MQTT, Modbus etc. and doesn't suffer from application specific OS or software requirements .
An analysis of the recorded data can be the foundation for insights into the challenges of distributed networks and later expand and adapt the tool for a new range of applications centering around the properties of distributed traffic replay. Use cases may revolve around scalability and reliability of networks, by giving users the opportunity to test their network, possibly introducing new traffic or emulating the network and its communication entirely without interfering with production processes.

The tool does not aim to replace existing IoT testing frameworks but rather enhances them. The tool has furthermore been made public\footnote{https://git.tu-berlin.de/markus.toll/iotreeplay}.

\textbf{Outline:} Section \ref{sec:background} and Section \ref{sec:related_work} provide an overview of the general subject as well as place this work within similar approaches. In Section \ref{sec:approach}, we present the design and implementation of \emph{IoTreePlay}.
Section \ref{sec:evaluation} contains the evaluation of replay experiments.
Finally, Section \ref{sec:conclusion} serves as a digression into future work and concludes the paper.

\section{Background}
\label{sec:background}
\emph{IoTreePlay} builds on the foundations of \textit{TCPLivePlay}. This section covers the functionality of \textit{TCPLivePlay} in its unmodified form.

\textit{TCPLivePlay} extends the \textit{TCPReplay} suite\footnote{https://tcpreplay.appneta.com} by allowing users to replay TCP communication from one host to another device.
It is written in C and uses the \textit{Free BSD pcap - Packet Capture Library}.
The tool replays a captured trace file. This is done by injecting packets directly at the NIC level, simulating a TCP connection without actually establishing one.
After providing destination IP, port and MAC-address, the capture file is read into an internal schedule. This schedule consists of all packets in the order in which they were recorded. In addition to marking every packet in the schedule as either a local or remote packet, showing whether it will be sent or should be received, the program initially sets up the schedule based on the recorded SEQs and ACKs. A random number is generated and used to offset all SEQ-numbers of local packets within the schedule. As a result, the SEQ-numbers now differ from their original but the difference between them remains the same. 

Furthermore, IP addresses and ports are rewritten to match the provided destination and IP of the source device, prompting for a recalculation of a all checksums.
In its unmodified version \textit{TCPLivePlay} assumes that the device on which it runs will initiate the connection and classifies all packets accordingly. After sending the first packet containing the SYN flag, it waits for the SYN/ACK response. When receiving said response the acknowledgment number of every local packet is updated so that it is based on the newly provided remote sequence number but is still consistent, in relative terms, to the original schedule. Equally all packets marked as remote packets are updated to represent the now expected remote sequence number. This is done so that a process can compare received packets to the corresponding packet from the capture file and check whether or not the connection runs as scheduled.

If expectations are met, the schedule proceeds to the next packet and the process is repeated until all scheduled packets are sent and received.
Since no organic connection is established but instead packets are just inserted at the network interface card, the operating system will usually respond with an RST flag when a TCP packet arrives at the chosen port, that is not currently listening for TCP connections. The installation guide covers how to stop this behaviour.
The program in its current version (4.4.1) contains semantic errors that hinder replay.

\section{Related Work}
\label{sec:related_work}

Multiple network replay tools were proposed by the scientific community already. Each with its unique focus on protocols, scenarios, or adjustment options.
The tool \textit{Mahimahi} developed by Netravali et al. in 2015 aims to combine records from HTTP-based applications as well as enable replay under emulated network conditions \cite{Mahimahi191577}. This enabled them to detect and improve on performance issues as well as introduce their own HTTP multiplexing protocol.

A team from Queensland University of Technology focused on packet capture and replay in their paper published in 2016. They were concerned with keeping the original timing of packets as precise as possible and not interfering with packet order or size in any way. This lead them to describe "[...]a hardware-software solution that builds on a special-purpose network interface card [...]" \cite[p.~9]{Forensics10.1145/2843043.2843047}.

In 2004 engineers at Google cooperated with a team from the University of California to create a tool for TCP tracing and replaying called \textit{Monkey See, Monkey Do}. It consists of two parts, the first is traffic capture from a server and the other offers replay capabilities. While it delivered in terms of accuracy in replay quality regarding session characteristics like delays, bandwidth, and packet loss they noted that the software might not be used as a generic TCP replay tool outside their specific use case. \cite{MonkeySeeMonkeyDo266375}.

\textit{TCPOpera} enables users to replay network traffic that is based around the functionalities of the \textit{TCPReplay} suite \cite{TCPOpera10.1007/11663812_13}. It focuses on adjusting network parameters like packet loss and delay and aims to replay the original file adjusted to these new parameters.

Our implementation follows a somewhat similar approach based around \textit{TCPLiveplay}. In contrast to the approaches described so far the focus lies on replaying TCP traffic in distributed networks. It will revolve around multiple nodes in the replay network while not requiring hardware adaptations or manipulating network characteristics to change output and provides a basis for future adaptions and additional features.

The IoT network emulator presented in~\cite{herrnleben2020iot} allows the tuning of network parameters inside an IoT framework. Users can emulate real-world traffic traces and vary certain network quality characteristics. However, no assistance is given to test application independent traffic scenarios or network-based fault injection. Our tool could be used to extend the proposed framework. 

IoT-Flock~\cite{ghazanfar2020iot} proposes an open-source framework to create and test IoT specific network traffic scenarios. The solution focuses on the application layer of the common IoT protocols MQTT and CoAP. IoT use-cases and devices can be created and subjected to generated normal and malicious IoT traffic. Goal of the work is to improve IoT network testing from a security perspective. While this approach is quite similar to ours, it is limited to certain protocols and does not allow the replaying of recorded network traffic. It is also not able to record and compare virtual and actual traffic.

\section{Design \& Implementation}
\label{sec:approach}

As previously described the goal of \emph{IoTreePlay} is to perform replay of TCP-based connections over a user-defined distributed network. This section provides descriptions of the problem as well as the theoretical approach and the underlying implementation.

\subsection{Problem}

Replaying multiple TCP connections within a network sets multiple challenges.

Firstly, TCP is a stateful protocol and so to repeat a data stream between two endpoints a new connection has to be established between them. Since the original connection has taken place at a different time if not between completely different devices, IP addresses of the data packets need to be adjusted. To remain accurate to the original packet count and size a new connection can not be established by traditional means. Resending the data would otherwise bear the risk of rearranging the data into a different amount of packets, deviating from the original.

Additionally, since TCP connections are bidirectional by design, both participants involved in a connection may need to send not only confirmation about the receival of data but may also send data themselves. To simulate this without access to the original data-producing processes both endpoints need to know which packet to send and to receive at what time before the connection can be established and used.

That is why before any of the actual replay processes can take place the distribution of all files containing said packets in their respective TCP streams must happen.

This not only requires the existence of multiple nodes in a network which provide the endpoints of the different connections but also a central controller that assigns roles to all participants and distributes the necessary information. A collection of nodes forms the IoT-network in which the replay will take place. All nodes need to be able to send TCP traffic but also be able to record said traffic.
The requirements for the controller are furthermore that it is capable of separating the original file into single connections. After the replay process finishes, all network recordings from each node must be retrieved.

\subsection{Design}

\begin{figure}[]
    \centering
    \includegraphics[width=\linewidth]{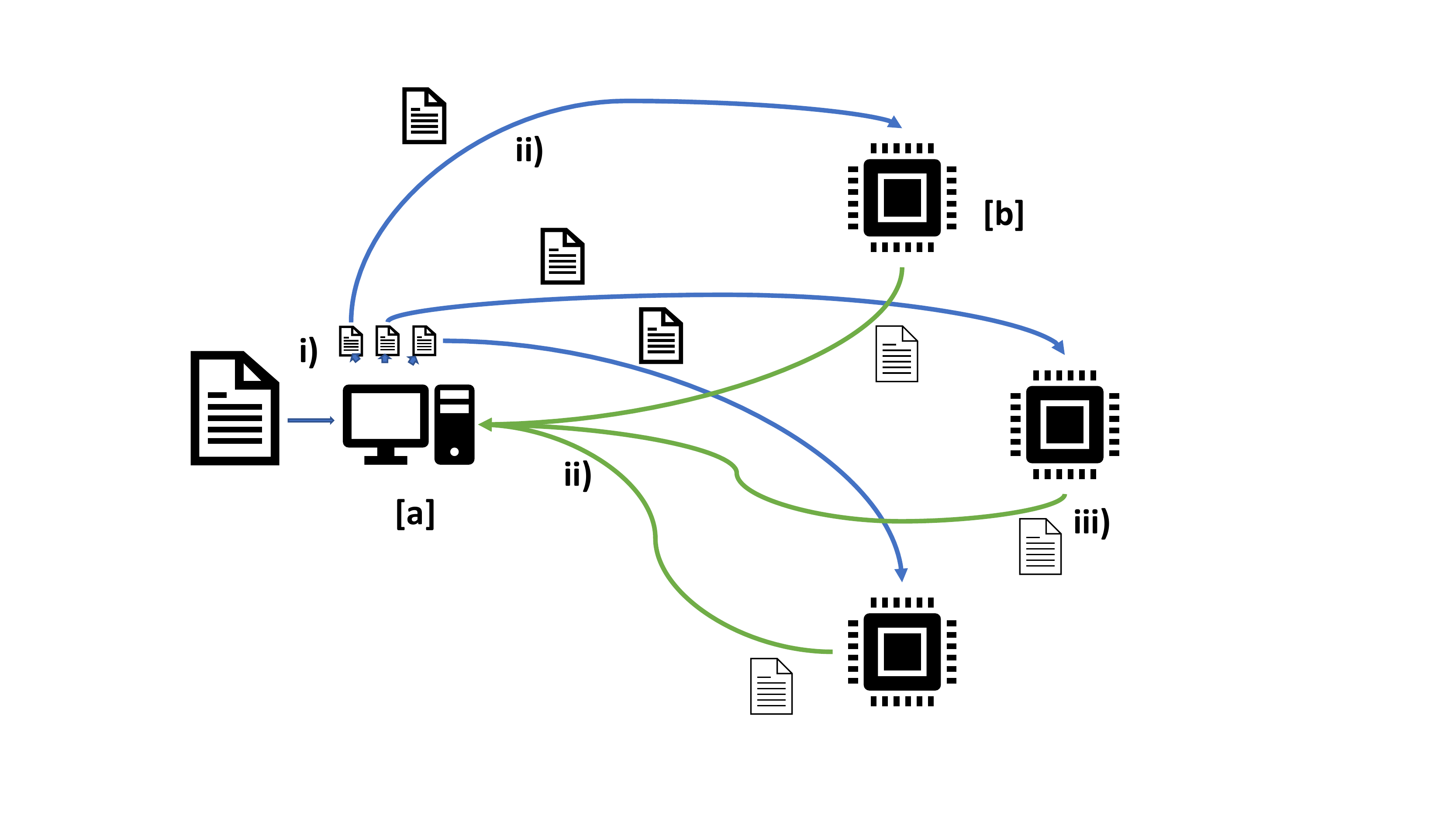}
    \caption{Depiction of Workflow}
    \label{fig:workflow}
\end{figure}

As can be seen in figure \ref{fig:workflow}, the work process consists of three distinct subprocesses involving two separate kinds of required devices.

The \textbf{Controller ([a])} initiates the replay process. It receives the user input and is responsible for splitting the original file into separate connections, as well as distributing these to every node and retrieving the recordings.

\textbf{Nodes ([b])} simulate devices within a distributed network. After receiving the workload they communicate among themselves thereby replaying the original traffic across the network. 

\subsection{Implementation}

At first, the network and its nodes have to be set up. This can either happen by using physical devices or emulating them virtually as long as the controller can access every node via SSH. For the purpose of this work, multiple Raspberry Pis were used.
The replay process is initiated by starting \textit{init.sh} and providing a capture file and a .txt file with SSH-Hosts as arguments.

\textbf{i) Splitting}
\textit{Tshark} separates the input file into files containing one connection. After completion of \textit{tshark}, the python code found in \textit{main.py} is run. All IPs from connections are read and stored. The user needs to provide a mapping for all IPs. Files with less than three packets (no handshake) are ignored. Additionally, a new port is set for each connection and every packet's source and destination port is overwritten with the new port number.

Files are then separated into directories for every network participant. Finally, every file is renamed following the same template so that file contents don't need to be analyzed remotely before starting \textit{TCPLivePlay}. The new name contains the original IPs, the stream number, the name of the input file, and lastly the port on which this connection should be replayed as well as a numerical value that marks the offset to the synchronized starting time. This offset is calculated based on the difference between the timestamps off a stream's first packet and the first stream's first packet.

\textbf{ii) Distribution and Retrieval}
The directory created for every device in the process described above is zipped and then transferred together with the \textit{remoteReplay.sh} file. Transfer takes place through \textit{Secure File Transfer Protocol} (SFTP). At the end of the replay process the host machine connects to every network participant again and retrieves the stored recordings. This then marks the final step of \textit{init.sh} and terminates the program.

\textbf{iii) Replay}
After file distribution the host uses \textit{parallel-ssh}\footnote{https://linux.die.net/man/1/pssh} to login to every device simultaneously and start the \textit{remoteReplay.sh} script. The current timestamp in seconds since the Epoch (01-01-1970) is provided as the first argument. This is the same way timestamps are stored in packets and is used to synchronize the start of replay on every device.
The script is responsible for starting the \textit{TCPLivePlay} instances and the capture process with \textit{tshark}.
After unzipping the transferred file, the script loops over every capture file and determines whether the connection is initiated locally or by another device. Additionally, the offset inscribed in the filename is used to adjust the synchronized start time. After all instances are started the capture of packets commences. The capture filter is set in a way that each device records all connections initiated by itself. After all \textit{TCPLivePlay} instances terminate, the capture is stopped and the resulting capture file is zipped again. This concludes the functionality of \textit{remoteReplay.sh}.

\textbf{TCPLivePlay Modifications}
\begin{itemize}
    \item \textbf{Errors}\newline
    The version of \textit{TCPLivePlay} that is included in \textit{TCPReplay V4.4.1} was not functioning. Firstly, it contained a timeout logic, originally intended to exit the program in case the remote host doesn't respond in time. This timeout was set to ten seconds and reset whenever a packet was sent or received. Unfortunately, the function handling incoming packets also had a timeout set to ten seconds. This second timeout lets the program wait ten seconds until an incoming packet can be processed upon arrival. Following that, \textit{TCPLivePlay} was never able to receive an answer to its first packet since it timeouts itself even though the packet would have been available. Additionally, this timeout would have delayed every packet by ten seconds completely ruining any attempt to accurately replay the input file.
    
    The second problem was that the schedule setup at the beginning of the program produced false values. This resulted in falsely calculated sequence or acknowledgment numbers for remote packets and resulted in classifying incoming packets as out-of-order even though they weren't.
    
    \item \textbf{Changes and Additions}\newline
    \textit{TCPLivePlay} automatically assumes that the device starting the program is the same that initiates the TCP connection from the input file. This required extensive changes in the setup structures of the program. Firstly, local IP couldn't by default be mapped onto the initiator of the connection. Following that, an "initiator" flag in the form of an integer is provided as the final argument when starting \textit{TCPLivePlay}. The value was determined by \textit{remoteReplay.sh} based on the input filename and the IP mapping provided by the user at the beginning of the process.
    
    A timestamp marking the beginning of the connection is provided at program start. This was calculated as described above by the sum of the synchronized start time for all instances and the individual offset for every connection. Furthermore, the delay between two packets needed to be implemented. \textit{TCPLivePlay} in its original form disregards timestamps and sends packets without delay. Delays between packets are calculated based on the timestamps in the IP header. These are then normalized so that the first packet is due at the beginning of the connection given by \textit{remoteReplay.sh}. Based on that whenever the program is at a point in the schedule where it's supposed to send a packet the newly calculated timestamp is checked against the current time and when a delay is necessary the inbuilt C usleep()-function with microsecond precision is called. The nanosleep()-function was not used because not all versions of .pcap-files support timestamps with nanosecond precision which would make the additional precision irrelevant. 

\end{itemize}

\section{Evaluation}
\label{sec:evaluation}

The main focus of the evaluation of \emph{IoTreePlay} was accuracy of replay in regards to the original timestamps of packets. This was achieved by comparing the original file with the recording after the replay, calculating the difference, herein referred to as deviation, between the recorded timestamps in the IP header. For every TCP connection replayed the deviation was first calculated for every packet and then aggregated over the whole connection. The aggregated metrics are minimum, maximum, average, median and standard deviation. With minimum and maximum being the earliest and latest a packet was recorded compared to its original. Multiple test runs were done. These included two-endpoint-connections with an transmission rate as well as multi-endpoint-connections with up to two hundred connections from publicly available repositories.

\begin{figure}
    \centering
    \includegraphics[width=\linewidth]{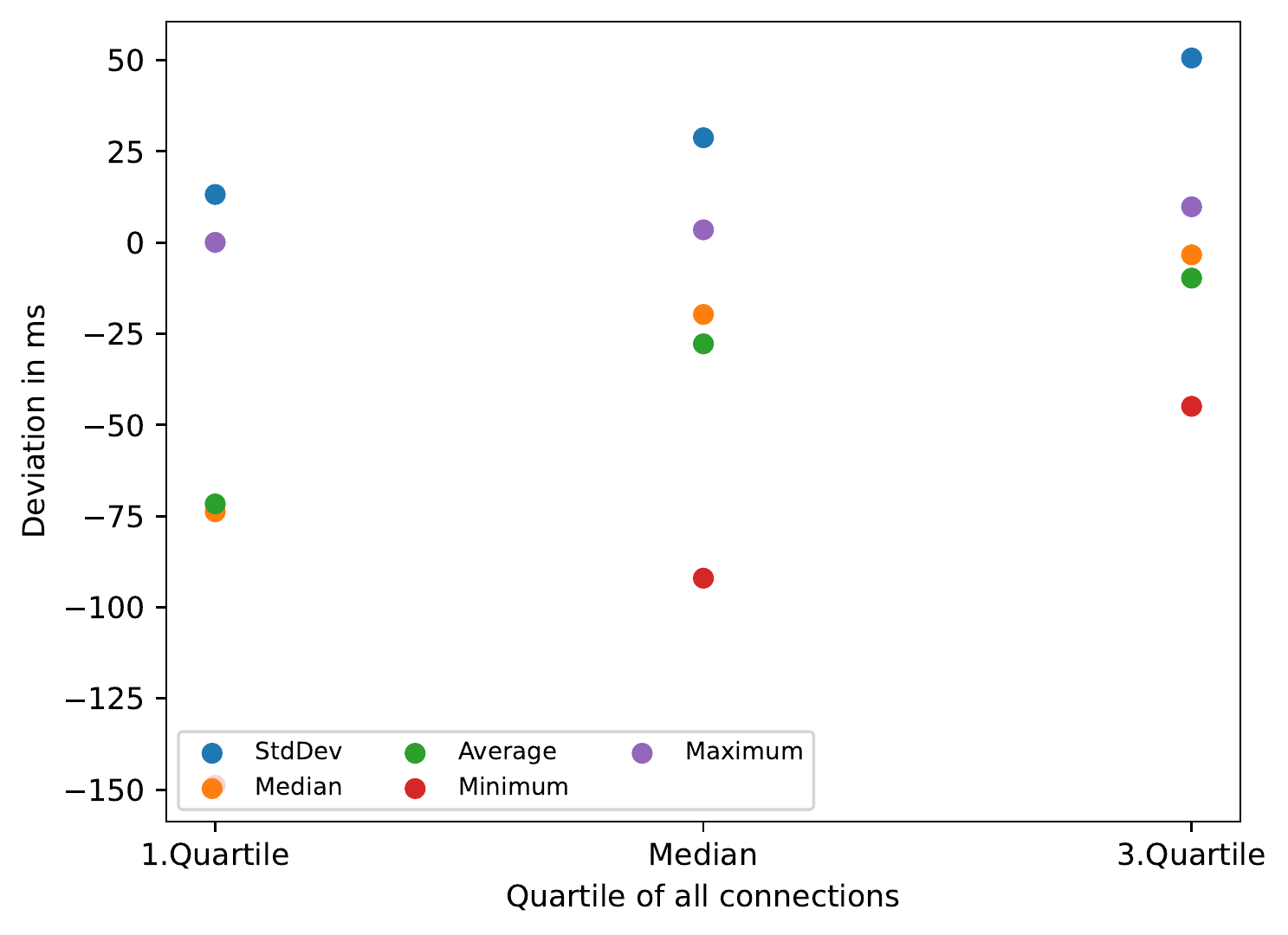}
    \caption{Metrics aggregated over all connections, in ms}
    \label{fig:deviationNormal}
\end{figure}

\begin{figure}
    \centering
    \includegraphics[width=\linewidth]{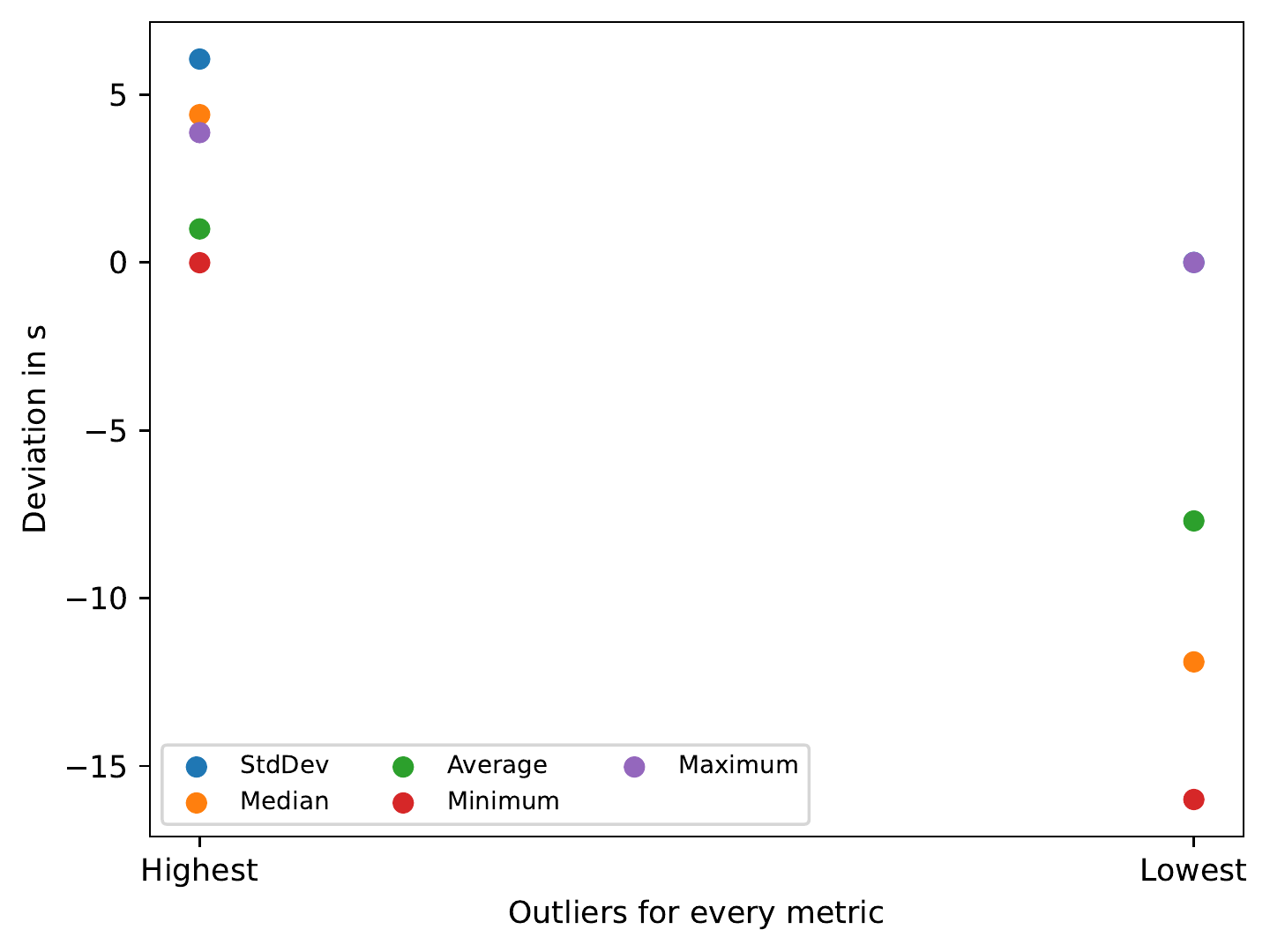}
    \caption{Outliers for every metric, in seconds}
    \label{fig:deviationOutliers}
\end{figure}

Figure \ref{fig:deviationNormal} shows the results aggregated over all connections. For more visibility outliers for every metric are visualized separately in figure \ref{fig:deviationOutliers}. Negative values represent that packets arrived later then they should, compared with the original packets timestamp. It can be observed hat even the earliest packets do not arrive significantly sooner than expected. The lines for average and median overlap heavily suggesting that within one connection the deviation was not heavily fluctuating around the average. Additionally, average and median, in the first quartile roughly halfway between maximum and minimum, find themselves closer to the maximum line later on. This suggests that fluctuation in packet deviation subsides in higher quartiles and is supported by the fact that the minimum line approaches zero. Overall, average and median deviation are in the area of 0-60 ms, potential reasons for which are discussed later.

The figure \ref{fig:deviationOutliers} displays the highest and lowest value recorded, for every metric. All values are in seconds and most of them belong to the same connection. One of the test runs contained MPTCP\footnote{https://www.multipath-tcp.org/} traffic. MPTCP is an effort to use the standard TCP interface but with multiple subflows simultaneously. This led to an increased throughput and a higher packet rate of more than a hundred packets per second. The replay was not capable of keeping up with these demands, marking a potential limitation in the maximum transmission capabilities of the current implementation. Nevertheless, the replay was not aborted prematurely suggesting that the limitation is performance based and not a general problem within the approach.

\begin{figure}
    \centering
    \includegraphics[width=\linewidth]{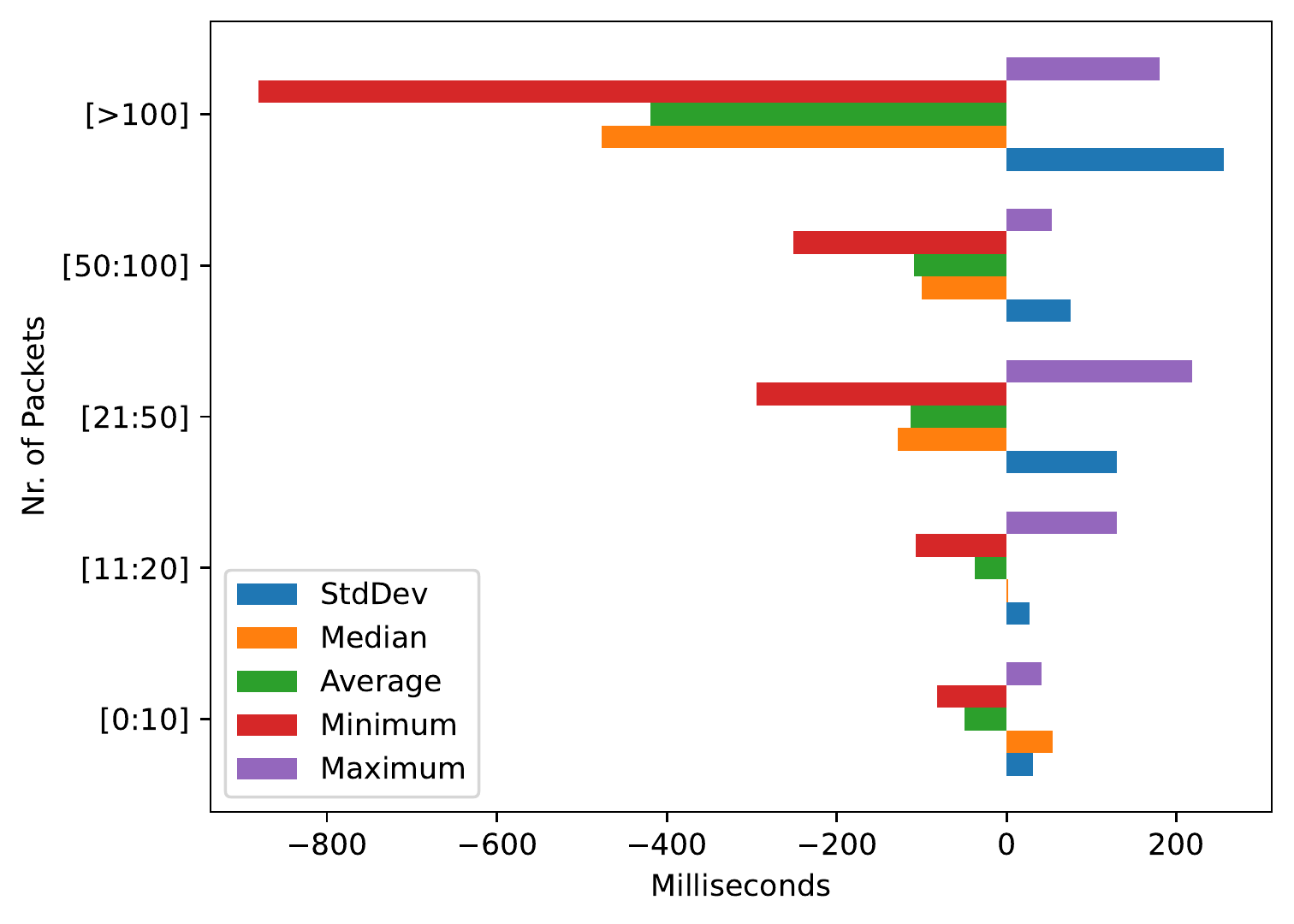}
    \caption{Metrics grouped by connection length (avg), in ms}
    \label{fig:groupedavg}
\end{figure}

\begin{figure}
    \centering
    \includegraphics[width=\linewidth]{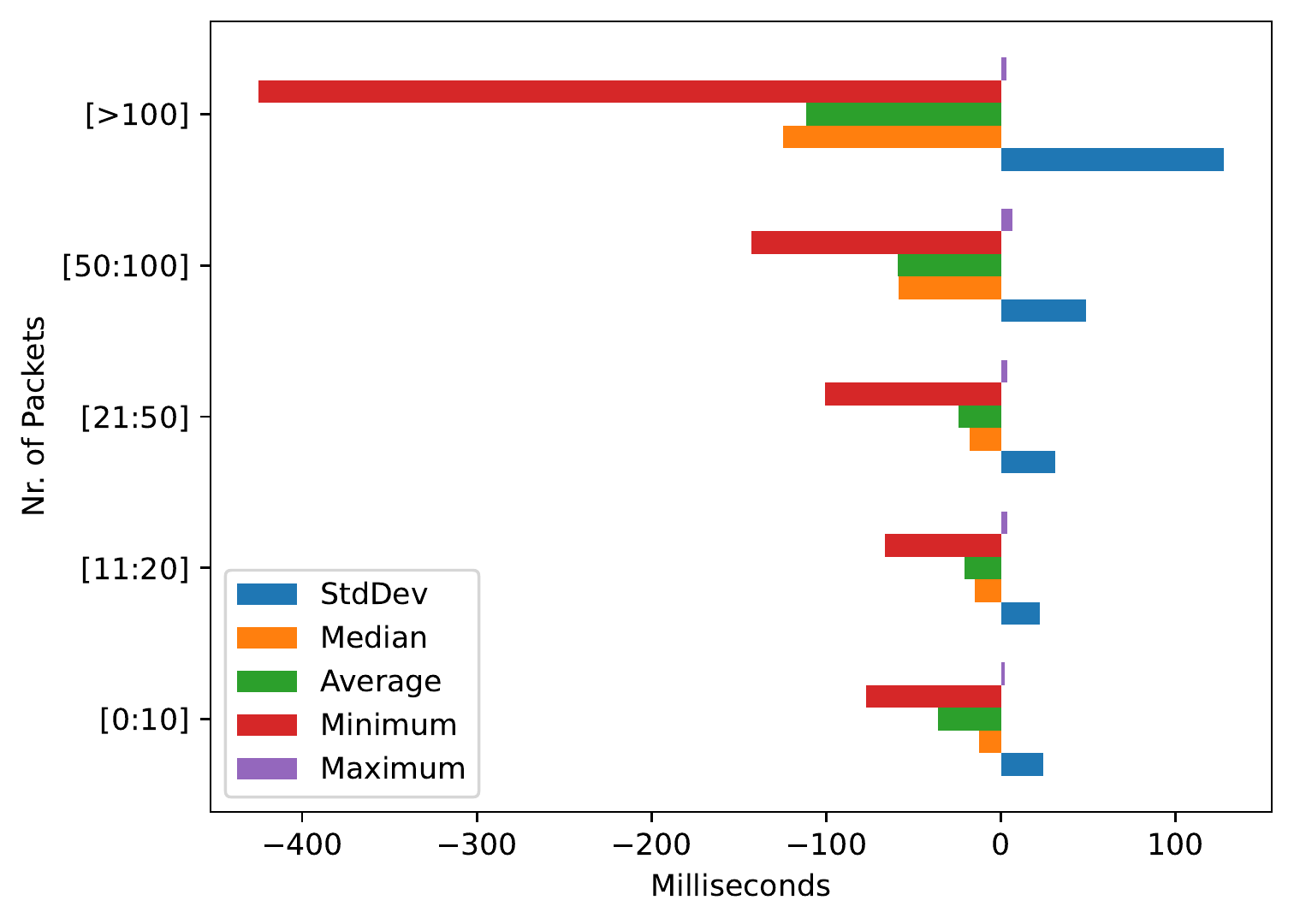}
    \caption{Metrics grouped by connection length (median), in ms}
    \label{fig:groupedmedian}
\end{figure}

Figures \ref{fig:groupedavg} and \ref{fig:groupedmedian} display every metrics average and median respectively, when aggregated by different connection lengths. First of all, it can be observed that when aggregating using the median all measurements are significantly lower. This shows, that outliers at the extreme influence all metrics heavily, which in turn suggests a minority of connections that are responsible for the majority of deviations. Both figures show that for all metrics except the maximum, which stays relatively unchanged, values increase with longer connections. With the minimum dimension changing the most and reaching the highest values in connections with more than one hundred packets. Connections reaching anywhere from zero to 50 packets perform roughly the same when looking at the median values. Notably, smaller connections seem to be the most accurate, suggesting that replay is accurate from the start and does not need time to stabilize. It should be mentioned that for connection lengths above 50 packets, significantly less connections were found within the test files. This likely explains some of the outliers at the upper end.

When putting together the results from above some insights can be gained. In general, the results show good accuracy for connections. Outliers at the extreme are, for the most part, due to technical limitations of the test setup and network structure and not caused by \emph{IoTreePlay} itself. Additionally, the test runs did not control for network parameters, like network delay between devices or computational delay at every device, that influence the result negatively.

While the test setup only included less then ten physical devices in the form of Raspbery Pis there was and likely will be no indication of scalability problems caused by the number of devices. Since all replayable data is split and distributed before the replay begins, replay capabilities and with that scalability is only determined by the processing power of a given node as well as the specifications of the network within which communication takes place. The only downside of larger replay files is that the initial splitting and distributing takes more time but it does not affect replay performance.
\textit{IoTreePlay} succesfully improves upon \textit{TCPLivePlay} by allowing a user to deploy multiple instances of \textit{TCPLivePlay} across a variety of nodes, emulating large numbers of IoT devices communicating within a network without the need for mock-up data processes and with no performance loss compared to a standard \textit{TCPLivePlay}, two-endpoint, replay.

A problem that occurred during test runs was that the replay success is heavily reliant on the structure of the input file. \textit{TCPLivePlay} includes some form of error correction for packets. Every arriving packet is not only checked against the expected schedule but also against packets that arrived previously. When connections already include duplicate packets, stemming from some form of packet loss in the original connection, \textit{TCPLivePlay} does not recognize them as such. When faced with a duplicate packets \textit{TCPLivePlay} thinks these come from packet loss during the replay. It reacts accordingly and retransmits previous packets, which in turn confuses the corresponding \textit{TCPLivePlay} instance on the other device, which does not expect this behaviour.
This leads to aborted connections or timeouts. This limitation can be somewhat mitigated by filtering out these kinds of packets before replay but that might not lead to the desired result.

Performance can be enhanced by streamlining the \textit{TCPLivePlay} implementation or rewriting it from the ground up. A more streamlined process with less error detection would likely increase throughput since less computational power and therefore time is needed managing arriving or departing packets. %

The above mentioned error correction functionality not only requires time but also leads to unwanted results. It is certainly possible to remove the error correction from \textit{TCPLivePlay} but that would effectively turn the replay into a UDP transmission with TCP packets. It then relies heavily on the underlying network connections stability. Another option is to add functionality that distinguishes between packet anomalies that are part of the schedule and the ones that are a result of the replay. Since all packets are fabricated by \textit{TCPLivePlay} and independent from the actual TCP protocol adaptions within the protocol header could provide a way to distinguish between said packets.

Additionally, right now delay between packets was determined by timestamps and enforced by sleep functions. When trying to optimize for minimal divergence the current approach is bound by the capabilities of the pcap.h and unistd.h libraries and their respective functionality.

\section{Conclusion \& Future Work}
\label{sec:conclusion}

\emph{IoTreePlay} delivers an easy to use and scalable software application that allows for TCP traffic replay within user-defined distributed networks. Good accuracy is achieved, especially within shorter connections.

The distributed approach for replaying traffic covers multiple use cases. When integrating distributed replay into an environment that allows for the configuration of network settings, the resilience of setups could be tested. This may involve a set of IoT devices producing network traffic. This traffic could then be replayed on a similar network with now changing parameters like network delay, simulating packet or connection loss, and others. The replay result would then provide insight into how stable the original network is and under what conditions it can still uphold acceptable connection parameters. This may be especially relevant in mobile applications, where network conditions change constantly.

Other fields of use may revolve around questions of scalability. Given an existing set of devices, could they handle additional or entirely new network load? This might be of interest for industrial digitization. Can existing devices still work properly, regarding overall performance but also regarding time-critical processes or might they be endangered by the introduction of communication processes beyond the already established? This could be tested by choosing network participants with specifications that are similar to the actual devices to be tested.

The replay capabilities may also be used for security applications. New traffic could be introduced into established networks trying to overload participants or to abuse open TCP ports and inject data into another device, either to infect or overload it.

Lastly, the functionalities can be expanded to include different protocols. As of now, only TCP traffic can be replayed. Depending on the application additional protocols may be desired. UDP would be an obvious candidate, with the \textit{TCPReplay} suite already providing possibilities for replay. But, communication standards don't have to be limited to IP-based protocols and may also include different forms of wired or wireless data transfer found in mobile or industrial distributed applications.

Given the simple setup of \emph{IoTreePlay}, it can be integrated into existing IoT testbeds to perform the functions mentioned above and more.
\balance
\bibliography{bibliography}

\end{document}